\documentclass[11pt]{article}
\usepackage{epsf,amssymb,latexsym,enumerate,cite,shadow,array,color}
\usepackage{slashed}
\usepackage{graphicx,wasysym}

\DeclareFontFamily{U}{rsf}{} \DeclareFontShape{U}{rsf}{m}{n}{
  <5> <6> rsfs5 <7> <8> <9> rsfs7 <10-> rsfs10}{}
\DeclareMathAlphabet\Scr{U}{rsf}{m}{n} \makeatletter
\@addtoreset{equation}{section} \makeatother

\def\be{\begin{equation}}
\def\ee{\end{equation}}
\def\ba{\begin{array}}
\def\ea{\end{array}}

\newcommand{\bea}{\begin{eqnarray}}
\newcommand{\eea}{\end{eqnarray}}

\def\K{K{\"a}hler}
\usepackage{ifpdf}
\ifx\pdfoutput\undefined
   \pdffalse
\else
   \pdfoutput=1
   \pdftrue
  \usepackage[pdftex]{hyperref}
\def\u0{{\underline 0}}

\def\url{{\underline {r+\ell}}}

\newcommand{\rf}[1]{(\ref{#1})}
\newcommand{\vp}{\varphi}

\newcommand{\vt}{\vartheta}

\begin{document}

\begin{titlepage}

\hskip 1cm

\vskip 3cm

\begin{center}
{\LARGE \textbf{   $\alpha $-Attractors:  Planck, LHC and Dark Energy}}

\

\

{\bf John Joseph M. Carrasco${}^{a}$},  {\bf Renata Kallosh${}^{\alpha}$},\,    {\bf  Andrei Linde${}^{\alpha}$} 
\vskip 0.5cm
{
${}^a$\small\sl\noindent Institut de Physique Th\'eorique,
CEA/DSM/IPhT, 
CEA-Saclay,\\
 91191 Gif-sur-Yvette, France
 \vskip 0.25cm
${}^\alpha$\small\sl\noindent SITP and Department of Physics, Stanford University, \\
Stanford, CA
94305 USA\\
}
\end{center}
\vskip 0.5 cm

\

\begin{abstract}

We develop four-parameter supergravity models of inflation and dark energy, constrained so that  ${\delta\rho\over \rho}$, $n_s$  and the cosmological constant $\Lambda $ take their known observable values, but where the mass of gravitino $m_{3/2}$ and the tensor-to-scalar ratio $r$ are free parameters.
We focus on generalized  cosmological $\alpha$-attractor models, with logarithmic \K\ potentials, a nilpotent goldstino and spontaneously broken supersymmetry at the de Sitter minimum. The future data on B-modes  will specify the parameter $\alpha$,  measuring the geometry of the \K\, manifold.
The  string landscape idea for dark energy  is supported  in these models  via an incomplete cancellation of the universal positive goldstino and negative gravitino contribution.   The scale of SUSY breaking  $M$  related to the mass of gravitino in our models is a  controllable parameter, independent on the scale of inflation, it will be constrained by  LHC data and future collider Energy-frontier experiments.
  
 \end{abstract}

\vspace{24pt}
\end{titlepage}

%\tableofcontents

\parskip 7pt

\section{Introduction}

\label{intro}

During the next few years we might expect some dramatic new information from B-mode experiments either detecting primordial gravity waves or establishing  a new upper bound  on $r$,  and from LHC discovery/non-discovery of low scale supersymmetry.  A  theoretical framework  to discuss both of these important factors in  cosmology and particle physics has been proposed  recently.  It is based on the construction of  new  models of chaotic inflation \cite{Linde:1983gd} in supergravity compatible with the  current cosmological data \cite{Planck:2015xua} as well as involving a controllable supersymmetry breaking at the minimum of the potential \cite{Kallosh:2014via,Dall'Agata:2014oka,Kallosh:2014hxa,Lahanas:2015jwa}. In this paper we will develop supergravity models of inflation motivated by either string theory or extended supergravity consderations, known as cosmological $\alpha$-attractors  \cite{Kallosh:2013hoa,Ferrara:2013rsa,Kallosh:2013yoa,Cecotti:2014ipa,Kallosh:2013tua,Galante:2014ifa,Kallosh:2015lwa,Kallosh:2015zsa,Carrasco:2015uma}. {\it Here we will enhance them with a 
 controllable supersymmetry breaking and cosmological constant at the minimum}. We find this to be  a  compelling framework for the discussion of the crucial new data on cosmology and particle physics expected during the next few years. Some models of this type were already discussed in \cite{Kallosh:2015lwa}. 
 
The paper is organized as follows.  We begin in Section~\ref{sect:Review} with a brief review of key vocabulary and features of these and related models with references to more in-depth treatments.
In Section~\ref{sect:Killing} we present the $\alpha$-attractor  supergravity models that make manifest an inflaton shift-symmetry by virtue of having the \K\ potential inflaton {\em independent} -- which we will refer to as Killing-adapted form.  Section~\ref{sect:Reconstruction} presents a  universal rule: given a bosonic inflationary potential of the form ${\cal F}^2 (\vp)$ one can  reconstruct the superpotential $W=\Big ( S +{1\over b} \Big) f(\Phi)$ for the  \K\, potentials described in Section~\ref{sect:Killing}. The resulting  models with $f'(\vp)={\cal F}(\vp)$ have a cosmological constant $\Lambda$ and an arbitrary SUSY breaking $M$ at the minimum. In Section~\ref{sect:genModels} we study more general class of models with $W= g(\vp) + S f((\vp)$ and the same \K\, potential. For these models it is also possible to get agreement with the Planck data as well as dark energy and SUSY breaking. Moreover, these models have nice properties with regard to initial conditions for inflation, analogous to the ones studied in \cite{Carrasco:2015rva} for models without SUSY breaking and dark energy.  We close in Section~\ref{sect:Discussion} with a summary of what we have accomplished.

 \section{Review}
 
 \label{sect:Review}
 
 \subsection{$\alpha$, and attraction}
 
 There is a key parameter $\alpha$ in these models, for which the \K\, potential $K= -3\alpha \ln (T+\bar T)$.  It describes the moduli space curvature \cite{Ferrara:2013rsa} given by ${\cal R}_K= - {2\over 3 \alpha}$.  Another, also geometric,  interpretation of this parameter is in terms of the Poincar\'e disk model of a hyperbolic geometry with the radius $\sqrt{3\alpha}$,   
illustrated by the Escher's picture Circle Limit IV~\cite{Kallosh:2015zsa,Carrasco:2015uma}. As clarified in these references, from the fundamental point of view, there are particularly interesting values of $\alpha$ depending on the original theory. From the maximal ${\cal N}=4$ superconformal theory, \cite{Cremmer:1977tt}, one would expect $\alpha=1/3$ with $r \approx10^{-3}$. This corresponds to the  unit radius Escher disk  \cite{Kallosh:2015zsa}, as well as a target of the future space mission for B-mode detection, as specified in CORE (Cosmic  ORigins Explorer). Some interesting simplifications occur for $\alpha = 1/9$, which corresponds to the  GL model \cite{Goncharov:1983mw,Linde:2014hfa}. From ${\cal N}=1$ superconformal theory \cite{Kallosh:2013hoa},  one would expect $\alpha=1$ with $r \approx 3 \times 10^{-3}$. Generic ${\cal N}=1$  supergravity allows any positive $\alpha$ and, therefore an arbitrary $r$, which has to be smaller than $0.11$ to agree with the current data.

\subsection{T and E model attractors, and observables}
A simple  class of  $\alpha$-attractor models, T-models,  have a potential  $V=\tanh^{2n} {\varphi\over \sqrt{6\alpha}}$ for the canonical inflaton field $\vp$. These models have
the following values of the cosmological observables \cite{Kallosh:2013hoa,Ferrara:2013rsa,Kallosh:2013yoa,Cecotti:2014ipa} for  $\alpha\lesssim O(10)$, where there is an attractor behavior and many models have the same $n$-independent predictions
 \be
  n_s  = 1-\frac{2}{N}\,, \qquad 
  r  =  \alpha \frac{12 }{N^2} \, , \qquad r\approx  3 \, \alpha \times 10^{-3} \  .
\label{aattr} 
\ee
 Once we increase $\alpha$ beyond $O(10)$,  expressions for $n_s$ and $r$ become somewhat different, see eqs. (5.2-5.4) in  \cite{Kallosh:2013yoa}. In particular, the value of $r$ can be increased significantly, all the way to the predictions of the $\vp^{2n}$ models. 
 \begin{figure}[htb]
%\vspace*{-2mm}%\hspace{-3mm}
\begin{center}
\includegraphics[width=9.3cm]{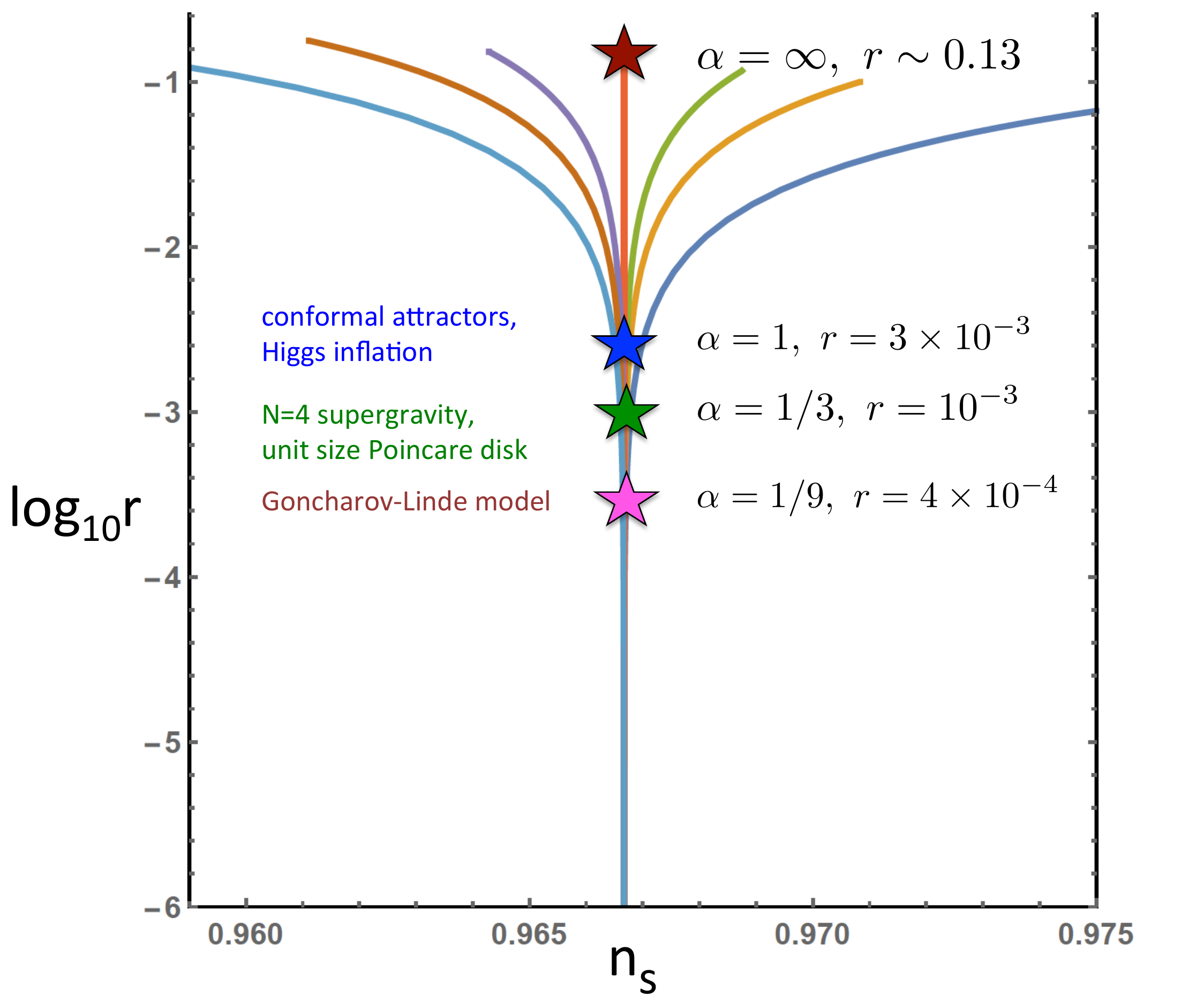}
\vspace*{-0.2cm}
\caption{\footnotesize  Examples of supergravity T- models with $r$-dependence in  logarithmic scale in $r$. For potentials
$V=\tanh^{2n} {\varphi\over \sqrt{6\alpha}}$, the predictions of these models  interpolate between the predictions of various polynomial models $\vp^{2n}$  at very large $\alpha$ and the vertical attractor line for $\alpha\leq O(10)$. When $\alpha \rightarrow \infty$ the models approach the ones with $\vp^{2n}$ potentials.  This attractor line beginning with the red star corresponds to the predictions of the simplest models $V=\tanh^{2n} {\varphi\over \sqrt{6\alpha}}$ with $n=1$.
}
\label{f2}
\end{center}
\vspace{-0.6cm}
\end{figure}
Even the simplest of these T-models are interesting phenomenologically for cosmology.  For these models  the parameter $\alpha$ can take any non-zero value; it describes the inverse curvature of the \K\ manifold \cite{Ferrara:2013rsa,Cecotti:2014ipa}. The cosmological predictions of these models, for various values of $\alpha$, are shown in Fig. 1.  As one can see, the line with $n=1$ begins at a point corresponding to the predictions of the simplest quadratic model ${m^{2}\over 2}\phi^{2}$ for $\alpha > 10^{3}$, and then, for smaller $\alpha$, it rapidly  cuts through the region most favored by the Planck data, towards the predictions of  the Higgs inflation model and conformal attractors $r \approx 0.003$ for $\alpha= 1$,  continues further down towards the prediction $r \approx  0.0003$ of the GL model \cite{Goncharov:1983mw,Linde:2014hfa} corresponding to $\alpha = 1/9$, and then the line goes even further, all the way down to $r \to 0$ in the limit $\alpha \to 0$.  This fact by itself is quite striking. 

\newpage

The simple  E-model attractors have a potential of the form  $V_{0} \Bigl(1- e^{-\sqrt {2\over 3\alpha} \varphi}\Bigr)^{2n}$. For $n = 1$, $\alpha = 1$ it gives the potential of the Starobinsky model, with the prediction $r \approx 0.003$. We will generalize both T-models as well as E-models, which both fit the data from Planck very well, to describe SUSY breaking and dark energy, at the minimum of the generalized potential.

\begin{figure}[h!]
\vspace*{3mm}
\centering
\includegraphics[width=9cm]{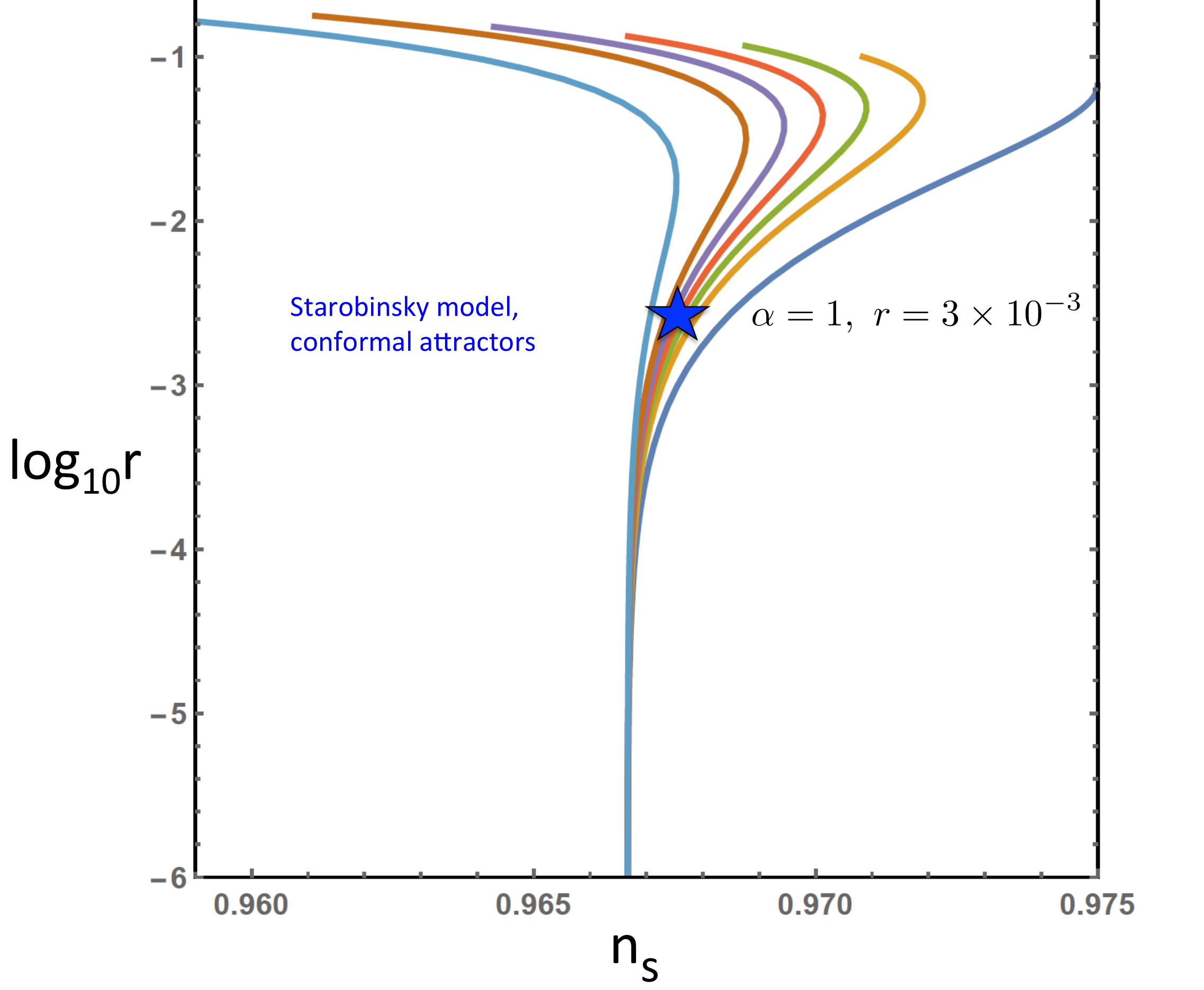}
\vspace*{-1mm}
\caption{\footnotesize  The cosmological observables $(n_s,r)$, in a logarithmic scale in $r$,   for  simple examples of E-models, with $V= (1- e^{-{ \sqrt {2\over 3 \alpha} \vp}})^{2n} $
with $n = (1/2, 3/4, 7/8, 1, 3/2, 2, 3)$  starting from the right, increasing to the left, with the vertical line for $n=1$ in the middle. When $\alpha \rightarrow \infty$ the models approach the ones with $\vp^{2n}$ potentials. The attractor line, common for all $n$, starts below $r\approx 10^{-3}$ and goes down, unlimited.}
\label{fig:simpleObservables}
\vspace{-.3cm}
\end{figure}

\

\subsection{Stabilizers}
In supergravity models of inflation, the task of SUSY breaking after inflation is often delegated to the so-called {\it hidden SUSY breaking sector}, requiring the addition of new superfields constrained to not participate in inflation. The scalars from such superfields have to be strongly stabilized, so as to not affect the inflation driven by the inflaton sector of the model.
In this paper we describe models of chaotic inflation  with the inflaton chiral superfield, and with a nilpotent superfield stabilizer.\footnote{The nilpotent multiplet describes the Volkov-Akulov fermionic goldstino multiplet with non-linearly realized spontaneously broken supersymmetry  \cite{Volkov:1973ix}. The relation to chiral nilpotent multiplets was studied in \cite{rocek}. In cosmology we use the recent implementation of nilpotent multiplets suggested in \cite{Komargodski:2009rz}. These nilpotent multiplets are deeply related to the physics of the D-branes \cite{Ferrara:2014kva,Kallosh:2014wsa}.}  This new approach to generic SUSY breaking was suggested recently in \cite{Kallosh:2014via} using generic  supergravity models including the inflaton multiplet as well as a nilpotent multiplet \cite{Ferrara:2014kva}.  Note that the non-inflaton goldstino multiplet plays 
an important  role for consistency  of inflation, including stabilization of the second scalar belonging to  the inflaton multiplet. This was explained in \cite{Kallosh:2010ug,Kallosh:2010xz} developing on the pioneering work \cite{Kawasaki:2000yn}. In these models   the glodstino multiplet was a `stabilizer' superfield and was a standard chiral superfield.

\subsection{Shift Symmetry and Z, T, and $\Phi$ variables}
 The inflationary models made with a shift-symmetric canonical \K\, potential, and controllable supersymmetry breaking have  been  studied in \cite{Kallosh:2014via,Dall'Agata:2014oka,Kallosh:2014hxa}.  The basic feature of all such models is as follows. At
the potential's minimum supersymmetry is  spontaneously broken. With the  simplest choice of the \K\, potential,  the models are given by
$
K= {1\over 2} (\Phi-\bar \Phi)^2 + S\bar S$, $ W= g (\Phi) + S f(\Phi)$, $ S^2(x, \theta)=0
$,
where the superpotential depends on two functions of the inflaton field $\Phi$. The difference with earlier models    \cite{Kallosh:2010ug,Kallosh:2010xz,Kawasaki:2000yn}, is  the presence of an $S$-independent function $g (\Phi)$ in $W$ and the requirement that $S$ is nilpotent.
 The mass of the gravitino at the minimum of the potential, $W=m_{3/2} =g(0)$, is non-vanishing in these new models,  and SUSY is broken in the goldstino direction with $D_SW =M \neq 0$. In \cite{Kallosh:2010ug,Kallosh:2010xz,Kawasaki:2000yn} the mass of the gravitino was vanishing. Typically the minimum of the potential is these models had an unbroken supersymmetry in Minkowski minima. But in new models in  \cite{Kallosh:2014via,Dall'Agata:2014oka,Kallosh:2014hxa} with $g(\Phi)\neq 0$ we find instead either de Sitter or Minkowski minima with spontaneously broken SUSY.

From the point of view of string theory and ${\cal N} \geq 2$ spontaneously broken supergravity, another class of \K\,  potentials, such as $K= -3\alpha \ln (T+\bar T)$,  is more interesting due to their geometric nature and symmetries. The same models in Poincar\'e disk variables  are given by $K= -3\alpha \ln (1-Z\bar Z)$. It is particularly important  that  these models have a boundary of the moduli space at 
\be
Z\bar Z \rightarrow 1 \,  , \qquad Z\rightarrow  \pm 1 \, ,  \qquad T \rightarrow   0\, ,  \qquad T^{-1} \rightarrow 0
\ee
where $T= {1+Z\over 1-Z}$, \, $T^{-1}= {1-Z\over 1+Z}$ \cite{Kallosh:2013hoa,Cecotti:2014ipa,Kallosh:2015zsa}. Inflation takes place near the boundary which 
leads to an  attractor behavior when many models lead to the same inflationary predictions. A simple way to explain it is to refer to a geometric nature of the kinetic terms of the form 
\be
 3 \alpha {\partial T \partial \bar T \over (T+\bar T)^2 }|_{T=\bar T=t}=  {3 \alpha \over 4}  \left ({\partial t  \over t }\right)^2= {3 \alpha \over 4}  \left ({\partial (t ^{-1}) \over t ^{-1}}\right)^2
\label{pole} \ee
The kinetic term has a pole behavior near $t^{-1}\rightarrow 0$, near the boundary of the moduli space $T^{-1} \rightarrow 0$. This explains why  the potentials can be changed without a change in cosmological observables and $r$ depends on the residue of the pole, i.e. on $\alpha$ \cite{Galante:2014ifa}. We may therefore change our potentials by small terms depending on  $t^{-1}$ without changing the observables during inflation.

 We study these models here.
They can use either the Poincar\'e disk variables $Z\bar Z < 1$ or the 
 half-plane variables $T+\bar T>0$.
 We will also use the set of variables discussed in \cite{Carrasco:2015rva}, where
 \be
T= e^{\sqrt{2\over 3\alpha} \Phi} \, , \qquad Z= \tanh {\Phi\over \sqrt{6 \alpha}} \ .
\label{Phi} \ee
In the context of our moduli space geometry the variables $\Phi$ represent the Killing adapted frame where the metric is inflaton independent. We will therefore call them Killing variables.

Our purpose here is to  generalize the models in  \cite{Kallosh:2013hoa,Ferrara:2013rsa,Kallosh:2013yoa,Cecotti:2014ipa}  to break ${\cal N} = 1$ SUSY spontaneously. 
The new models with $S^2(x, \theta)=0$, which are compatible with established cosmological data and designed to be compatible with the future data on $r$ and $m_{3/2}$ will depend on four parameters: $\alpha$, describing the \K\, geometry,  $M$, defining the scale of SUSY breaking by goldstino $D_SW= M$, and $\mu$, related to scale of inflationary energy and $b$. The role of $b$ is the following:
at the minimum  
\be
V= \Big (b^2-3 \Big ) {M^2\over b^2} \, ,  \qquad \Rightarrow \qquad     b^2=3\, , \qquad  V=0 \ .
\label{V1}\ee
It shows that in ${\cal N}=1$ d=4 supergravity with a nilpotent  goldstino multiplet {\it generic de Sitter minima require a universal condition that the goldstino energy $M^2$ exceeds the negative gravitino contribution  to energy}  where $ m_{3/2}^2=  {M^2\over b^2}$.
\be
V=  M^2-  3 m_{3/2}^2 >0 \ .
\label{V2}\ee
We keep here generic values of the parameter $b^2 > 3$ which allow generic de Sitter vacua of the string landscape type, including the case 
\be
\Lambda =M^2-  3 m_{3/2}^2= \Big (1-{3\over b^2} \Big ) M^2\sim 10^{-120} \ .
\label{V3}\ee

\section{ Killing-adapted $\alpha$-attractor supergravity models.}
\label{sect:Killing}
We study here the following  ${\cal N} = 1$ supergravity models, which can be described  in disk geometry coordinates  of the moduli space $Z$, 
\be
 K= -3  \alpha \log \Big (1- Z\bar Z  \Big ) +S\bar S\, , \qquad S^2(x, \theta)=0\, ,  \qquad W= \tilde A(Z)  + S \tilde B(Z)\,   \ .
\label{Kdisk}\ee
The geometry has the $SU(1,1)$ symmetry
\be
ds^2= K_{Z\bar Z} dZ d\bar Z= -3\alpha {dZ d\bar Z\over (1-Z\bar Z)^2} \ .
\label{Dgeom}\ee
Alternatively, we can use the half-plane coordinates $T$ 
 \be
K= -3\, \alpha  \log \left(T + \bar T    \right) + S\bar S\, ,  \qquad S^2(x, \theta)=0\, ,  \qquad W= \tilde G(T)+ S \tilde F(T)\ .
\label{Khalf}\ee
The geometry has an $SL(2, \mathbb{R})$ symmetry
\be
ds^2= K_{T\bar T} dT d\bar T= -3\alpha {dT d\bar T\over (T+\bar T)^2} \ .
\label{HPgeom}\ee
In both cases, at $S=0$ the geometry is associated with the Poincare disk or half plane  geometry where $3\alpha= R_{E}^2$ corresponds to the radius square of the Escher disk \cite{Kallosh:2015zsa}. 

We will now perform a \K\, transformation \cite{Carrasco:2015uma,Carrasco:2015rva} so that our new \K\, potential is  inflaton shift-symmetric. First we use the original disk and half-plane variables and redefine the \K\, and superpotentials as follows
\be
K= -{3\over 2}   \alpha \log \left[{(1- Z\bar Z)^2\over (1-Z^2) (1-\overline Z^2)}  \right] +S\bar S\, , \qquad S^2(x, \theta)=0\, ,  \qquad W= A(Z)  + S B(Z)\,   \ .
\label{KdiskNew}\ee
where 
\be
 A(Z) + S B(Z) =(1-Z^2)^{-3\alpha /2} ( \tilde A(Z)  + S \tilde B(Z)) \ .
\ee
In half-plane case
 \be
K= -{3\over 2}\, \alpha  \log \left[ {(T + \bar T )^2 \over  4 T \bar T}   \right] + S\bar S\, ,  \qquad S^2(x, \theta)=0\, ,  \qquad W= G(T)+ S F(T)\ .
\label{KhalfNew}\ee
where
\be
G(T) + S F(T) = T^{-3\alpha /2} \bigl(\tilde G(T)+ S \tilde F(T)\bigr) \ .
\ee
Since we have performed a \K\, transform of the type
\be
K\rightarrow K + {3\alpha \over 2} \log [(1-Z^2)  (1-\bar Z^2)], \qquad W\rightarrow (1-Z^2)^{-3\alpha /2} W\, \qquad \overline W\rightarrow  (1-\bar Z^2) ^{-3\alpha /2} \overline W \ .
\ee  
\be
K\rightarrow K + {3\alpha \over 2} \log  [4 T  \bar T], \qquad W\rightarrow T^{-3\alpha /2} W\, \qquad \overline W\rightarrow \bar T^{-3\alpha /2} \overline W \ .
\ee  
the geometry did not change, it is still given by \rf{Dgeom} and \rf{HPgeom}, respectively.

Our next step is to switch to moduli space coordinates \rf{Phi} where the metric is manifestly inflaton-independent. {\it The choice of coordinates   
$Z= \tanh {\Phi\over 6 \alpha}$ and $
T= e^{\sqrt{2\over 3\alpha} \Phi} $ in the disk/half-plane geometry 
 corresponds to a Killing-adapted choice of coordinates where the metric does not depend on $\vp = {\rm Re} \, \Phi$}. We find that in these coordinates with Killing variables $\Phi= \vp +i \vt$
 \be
 K= -3\alpha \log \Big [\cosh {\Phi-\bar \Phi \over \sqrt{6\alpha}} \Big] + S \bar{S} \ .
\ee
and 
\be
ds^2=  -3\alpha {dZ d\bar Z\over (1-Z\bar Z)^2}= -3\alpha {dT d\bar T\over (T+\bar T)^2}= {\partial \Phi \partial \bar \Phi\over 2 \cos^2\Big (\sqrt{2\over 3\alpha}\, {\rm Im} \Phi\Big )}  .
\ee
The superpotential is now
\be
 W= A \Big (\tanh
 {\Phi \over \sqrt {6\alpha}}
 \Big)  + S\, B \Big (\tanh
 {\Phi \over \sqrt {6\alpha}}\Big ) = G\Big (e^{ \sqrt{2\over 3\alpha}  \Phi }\Big) + S F\Big ( e^{ \sqrt{2\over 3\alpha}  \Phi }\Big)
\,   \ .
\ee
Note that in our models $\vartheta=0$ during inflation and therefore the new holomorphic variable $\Phi$ during inflation becomes a real canonical variable
$\vp$. This is also easy to see from  the kinetic terms in these variables, which are conformal to flat,
\be
 ds^2=    {d\vp^2 + d\vartheta^2 \over 2\cos^{2}\sqrt{2  \over 3\alpha} \vartheta } \ .
\label{JJ}\ee
At $\vt=0$ they are both canonical
$
 ds^2|_{\vt=0} =    {d\vp^2 + d\vartheta^2 \over 2 }
 $. 
 Thus, we will work with  $\alpha$-attractor models \rf{Kdisk}, \rf{Khalf} in
 the form
\be
 K= -3\alpha \log \Big [\cosh {\Phi-\bar \Phi \over \sqrt{6\alpha}} \Big] + S \bar{S}\, , \quad   W= G\Big (e^{ \sqrt{2\over 3\alpha}  \Phi }\Big) + S F\Big ( e^{ \sqrt{2\over 3\alpha}  \Phi }\Big).
\label{new} \ee
Here one should keep in mind that our original half-plane variable $T$ is related to $\Phi$ as follows, $T= e^{ \sqrt{2\over 3\alpha}  \Phi }$.
We will  use the following notation
\be
G\Big (e^{ \sqrt{2\over 3\alpha}  \Phi }\Big)\equiv  g(\Phi)\, ,  \qquad F\Big ( e^{ \sqrt{2\over 3\alpha}  \Phi }\Big)\equiv f(\Phi) \ .
\ee

\noindent { \it To summarize, in Killing variables the $\alpha$-attractor supergravity models are}
\be 
 K= -3\alpha \log \Big [\cosh {\Phi-\bar \Phi \over \sqrt{6\alpha}} \Big] + S \bar{S}\, , \quad   W= g(  \Phi ) + S f(  \Phi ) \ .
\label{newR} \ee
We find that  the potential at $\Phi=\bar \Phi$  and at $S=0$ is given by 
\be
V_{\rm total}=  2 |g'(\vp)|^2 - 3  |g(\vp)|^2 +  |f(\vp)|^2 \ ,
\ee
since
the \K\, covariant derivatives are the same as simple derivatives 
\be
D_\Phi W = \partial _\Phi W=g'(\Phi) \, ,  \qquad D_S W= \partial_S W= f(\Phi) \ ,
\ee
and at $\Phi=\bar \Phi$,  $S=0$,  $K=0$ and the inverse kinetic terms  $K^{S\bar S}=1$ and $K^{\Phi\bar \Phi}=2$.

\section{ Reconstruction models of inflation with  SUSY breaking and de Sitter exit }
\label{sect:Reconstruction}
In the form \rf{newR} our $\alpha$-attractor models can be used to provide a de Sitter exit from inflation as well as supersymmetry breaking at the minimum of the potential, without changing any of the advantages in describing inflation. One of the simplest possibilities for such models is to require that 
\be
g(\Phi) ={1\over b} f(\Phi)  \ ,
\label{W}\ee
\be
 K= -3\alpha \log \Big [\cosh {\Phi-\bar \Phi \over \sqrt{6\alpha}} \Big] + S \bar{S}\, , \quad   W= \Big (S+{1\over b} \Big) f(   \Phi ) \ . \label{new1} \ee
In Killing  variables we find that at  $\Phi=\bar \Phi$  and at $S=0$
\be
D_\Phi W = \partial _\Phi W={1\over b}  f'(\Phi) \, ,  \qquad D_S W= \partial_S W= f(\Phi) \ .
\ee
The expression for the potential at $\Phi-\bar \Phi=S=0$ is now very simple and is given by
\be
V= \Big (1-{3\over b^2} \Big ) |f (\vp) |^2 + {2\over b^2} | f'(\vp) |^2 \ .
\label {V}\ee
Assume that at the minimum of the potential at $\Phi=0$ 
\be
f (0)= D_S W= M \neq 0\, , \qquad f'(0) = b \, D_\Phi W= 0 \ .
\ee 
This means that at the minimum supersymmetry is broken only in the direction of the nilpotent superfield $S$ and unbroken in the inflaton direction, since $b\neq 0$.

We take   $b^2> 3$. This provides an opportunity to have de Sitter vacua with positive cosmological constant $\Lambda$  in our inflationary models so that
\be
V|_{\Phi=0} = \Lambda\, , \qquad         \Lambda\equiv  \Big (1-{3\over b^2} \Big ) M^2\, ,  \qquad b^2 = {3\over 1-{\Lambda\over M^2}} \ .
\ee
 The cosmological constant is extremely small, $\Lambda \sim 10^{{-120}}$, so we would like to make a choice of $f$ in \rf{W} such that the inflationary potential is presented by the second term in \rf{V}. In such case, with account of $\vt=0$ condition we can use the reconstruction method analogous to the one in \cite{Dall'Agata:2014oka}, where it was applied to canonical shift symmetric \K\, potentials with Minkowski vacua. We will show here how to generalize it for de Sitter exit from inflation and our logarithmic \K\, potentials.

If the potential during inflation is expected to be given by the function
\be 
V(\vp) = {\cal F}^2(\vp) \ .
\label{potential}\ee
we have to take
\be
\partial _\vp f(  \vp )= {b\over \sqrt{2}}\,  {\cal F}(\vp) \ ,
\ee
 and 
 \be
 f( \vp ) =  {b\over \sqrt{2}}\, \int  {\cal F}(\vp) \qquad   f(  \vp )|_{\vp=0} =M \ .
 \ee
In these models the value of the superpotential at the minimum defines the mass of gravitino as follows
\be
W_{\rm min}= {f\over b}|_{\Phi=0}  = {M\over b}= {M\over \sqrt 3} \Big (1-{\Lambda\over M^2}\Big )^{1/2}= m_{3/2} \ ,
\ee
where
$
\Lambda= M^2-3 m_{3/2}^2
$. The total potential at $\vt=0$ is therefore given by
\be
V^{\rm total}= \Lambda   {|f (\vp)|^2 \over M^2} +|{\cal F}(\vp)|^2 \ ,
\label{total}\ee
with 
\be
V^{\rm total}_{\rm min}= \Lambda   = M^2-3 m_{3/2}^2 \ .
\label{L}\ee
To get from the supergravity model \rf{new1} to the Planck, LHC, dark energy potential \rf{total} requires stabilization of the field $\vt$ at $\vt = 0$. We have checked that for all values of $\alpha$ during inflation, up to slow roll parameters, the main contribution to the mass to Hubble ratio is of the form
\be
{m^2_\vt
\over H^2} \approx   6 {|f |^2\over |f'|^2} \gg 1 .
\label{6a}\ee
Here  the mass of $\vt$ is defined with a proper account taken of the non-trivial kinetic term. Equation  \rf{6a} implies that $\vt$ quickly reaches its minimum at $\vt=0$ at the bottom of the de Sitter valley, and inflation proceeds due to a slow evolution of $\vp$. 
However, near the minimum of the potential, where the slow roll parameters are not small, a more careful  evaluation of the mass of $\vt$ has to be performed. We will do it in examples below.

\subsection{ The simplest  T-model  with broken SUSY  and dS exit}
We would like to have the inflationary  part of the the potential to be
\be\label{nobreakFanned}
V_{\rm infl}(\vp) = \alpha \,\mu^{2}\tanh^{2} {\vp\over \sqrt {6\alpha}} \ .
\ee
This means that 
\be
{\cal F}= \sqrt \alpha \mu   \, \tanh {\vp\over \sqrt {6\alpha}}
\ee
 and
   \be
   f(\vp)= \sqrt  3\, \alpha\, \mu \, b\, \log \Big [\cosh  {\vp\over \sqrt {6\alpha}}\Big]
 +M \ .
 \ee
At $\vp=0$ one has $f(\vp)=M$. A complete supergravity version of the model is
 \be
 K= -3\alpha\, \log \Big [\cosh {\Phi-\bar \Phi \over \sqrt{6\alpha}} \Big] + S \bar{S}\,   , \qquad W=  \Big (S+{1\over b}\Big ) \Big [\sqrt  3  \, \alpha \, \mu \, b\, \log \Big[ \cosh {\Phi\over \sqrt {6\alpha}}\Big] 
+M \Big ] \, .
\label{Tsugra}\ee
The total potential has a part proportional to the cosmological constant $\Lambda$ as well as the second part describing inflation:
\be
V_{\rm total}= \Lambda    {|f (\vp)|^2 \over M^2} + \alpha \,\mu^{2}\tanh^{2} {\vp\over \sqrt {6\alpha}} \ .
\label{Tpot}\ee

The issue of the $\vt$ field stabilization which is required to get  from \rf{Tsugra} to \rf{Tpot} presents an example of the general case. We find that during inflation ${m^2_\vt
\over H^2}$ is positive and large, $\vt$ quickly reaches 0.  
However, near the minimum of the potential,  the evaluation of the mass of $\vt$ shows that it is positive under condition that $\alpha  \gtrsim 0.2$. Thus for $r\gtrsim 10^{-3}$ the model is safe without any stabilization terms even at the de Sitter minimum. For smaller $\alpha$ the bisectional curvature term has to be  added to the \K\, potential, to stabilize $\vt$. It is given by an expression in disk variables of the form $A(Z, \bar Z) S\bar S (Z-\bar Z)^2$.

\begin{figure}[htb]
%\vspace*{-2mm}%\hspace{-3mm}
\begin{center}
\includegraphics[width=8cm]{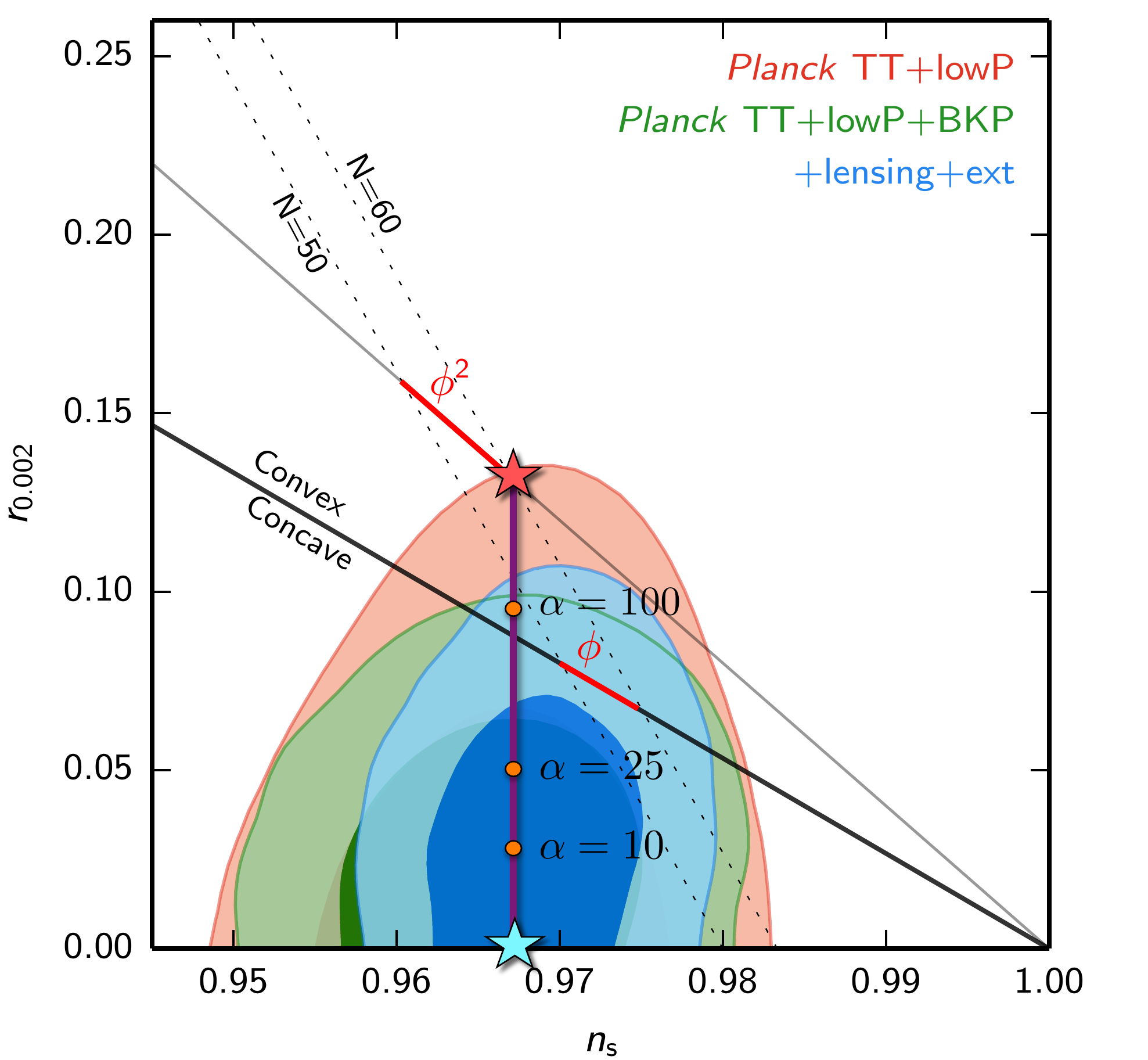}
\vspace*{-0.2cm}
\caption{\footnotesize  Cosmological predictions of the simplest T-model \rf{Tpot} with SUSY breaking and a  non-vanishing cosmological constant $ \Lambda \sim 10^{{-120}}$.
}
\label{f2a}
\end{center}
\vspace{-0.6cm}
\end{figure}

The cosmological predictions of this model are represented by the straight vertical line in Fig. 1. A more direct comparison with the Planck results is provided by a figure presented in \cite{Kallosh:2015lwa}, which we reproduce here as Fig. \ref{f2a}.

Note that this model in disk variables and in a different \K\, frame was already presented in eqs. (3.20) and (3.21) in  \cite{Kallosh:2015lwa}. An interesting property of the model \rf{Tpot} is that the amplitude of scalar perturbations does not depend on $\alpha$ and is determined only by $\mu \approx 10^{-5}$.

\subsection{ The simplest  E-model  with broken  SUSY  and dS exit}
We are looking at the inflationary  $\alpha$ model with 
\be
V_{\rm infl} = m^2 
\Big(1- e^{-\sqrt {2\over 3\alpha} \varphi}\Big)^2 \ .
\ee
This  means that 
\be
{\cal F}= m  \Big(1- e^{-\sqrt {2\over 3\alpha} \varphi}\Big)
\ee
 and
   \be
   f(\vp)= {m b\over \sqrt 2} \Big ( \vp + \sqrt {3\alpha \over 2} e^{-\sqrt {2\over 3\alpha} \vp} -1 \Big )+M \ .
   \ee
At $\vp=0$ one has $f(\vp)=M$.

Thus our complete model is
 \be
 K= -3\alpha \log \Big [\cosh {\Phi-\bar \Phi \over \sqrt{6\alpha}} \Big] + S \bar{S}\,   , \qquad W=  \Big (S+{1\over b}\Big ) \Big [{mb\over \sqrt 2}  ( \Phi + \sqrt {3\alpha \over 2} e^{-\sqrt {2\over 3\alpha} \Phi} -1)+M \Big ] \, .
\label{Esugra}\ee
The total potential has a part proportional to the cosmological constant $\Lambda$ as well as the second part describing inflation:
\be
V_{\rm total}= \Lambda   {|f(\vp) )|^2\over M^2} +m^2 
\Big(1- e^{-\sqrt {2\over 3\alpha} \varphi}\Big)^2 \, .
\label{Epot}\ee
The issue of the $\vt$ field stabilization which is required to get  from \rf{Esugra} to  \rf{Epot} has been studied separately and again confirms the general case as discussed below eq. \rf{L} concerning inflationary part. And again 
 near the minimum of the potential,  the evaluation of the mass of $\vt$ shows that it is positive under condition that $\alpha >0.2$.  For smaller values of $\alpha$, the bisectional curvature term has to be  added to the \K\, potential, to stabilize $\vt$. It is of the form $A(Z, \bar Z) S\bar S (Z-\bar Z)^2$
in disk variables.
This model for $\alpha=1$ in half-plane variables in case of $\Lambda=0$ was proposed in \cite{Lahanas:2015jwa} in eqs. (28), (37). For the generic case of $\alpha\neq 1$ a related model was given in eqs. (4.23),  (4.24) in \cite{Kallosh:2015lwa}.

More general models can be constructed following the rules  for this class of models proposed above in eqs. \rf{potential} - \rf{total}.

\section{General models of inflation with SUSY breaking and  dark energy} 
\label{sect:genModels}

We have learned above how to build supergravity models by reconstructing superpotentials to produce a given choice of the bosonic inflationary potential $V(\vp) = {\cal F}^2(\vp)$ with our  logarithmic \K\, potential $K= -3\alpha\, \log \Big [\cosh {\Phi-\bar \Phi \over \sqrt{6\alpha}} \Big] + S \bar{S}$ in Killing variables. The exact answer for $ W= g(  \Phi ) + S f(  \Phi )$ can be obtained under condition $g(\Phi) ={1\over b} f(\Phi)$ 
 and requires simply an integration so that $f( \vp )$ is reconstructed by integration  $f( \vp )=  {b\over \sqrt{2}}\, \int  {\cal F}(\vp)$.    Obviously this can be carried out in any variables as long as one takes care of the \K\ measure relating the variables used to the functional form of the canonical variables, but it is particularly transparent in Killing-adapted variables as the measure is unity.

Instead of the reconstructing strategy we may start with our models in \rf{newR} with  superpotentials of the form
\be
W= g(\Phi) + S f(\Phi)
\ee
without a constraint that $g(\Phi) ={1\over b} f(\Phi)$. In such case the potentials are given by 
$
V_{\rm total}=  2 |g'(\vp)|^2 - 3  |g(\vp)|^2 +  |f(\vp)|^2
$.

Near the minimum of the potential one has to check that we still satisfy the requirements that $D_S W= M \neq 0$ and  $ D_\Phi W= 0$ to preserve the nice de Sitter exit properties with SUSY breaking as described in eq. \rf{V1}.
In these models we  end up with more complicated bosonic potentials describing some combination of our $\alpha$-attractor models. However, these models are still capable to fit the cosmological observables as well as providing the level of SUSY breaking in  dS vacua with a controllable  gravitino mass. Some examples of these models were given in \cite{Kallosh:2015lwa},  in eqs. (2.4), (3.15) and (2.7), (3.17).
\begin{figure}[htb]
\vspace*{-2mm}%\hspace{-3mm}
\begin{center}
\includegraphics[width=11cm]{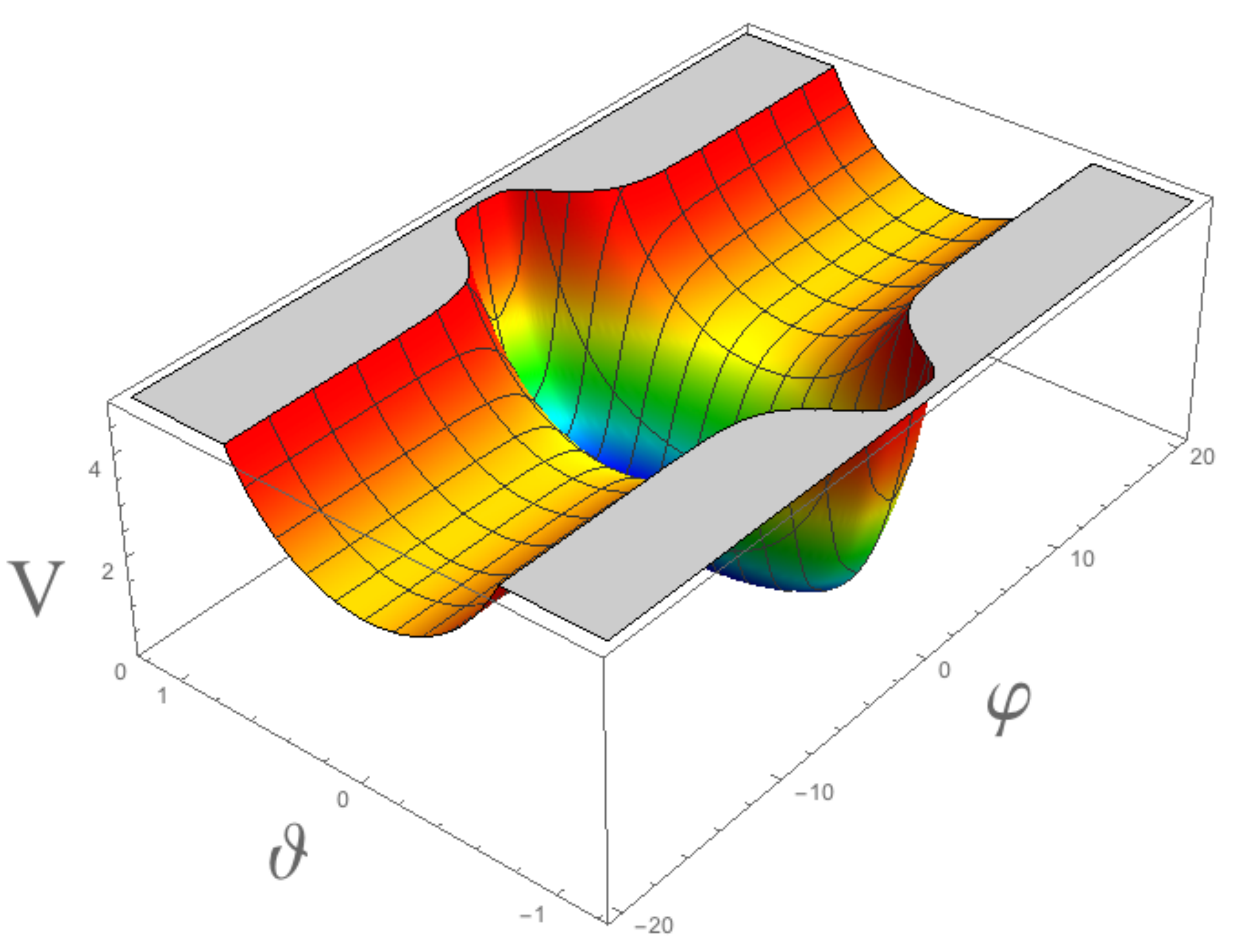}
\vspace*{-0.2cm}
\caption{\footnotesize  The potential for the supergravity model in eq. \rf{newRexample} as a function of $ \vp$ and $\vt$.
It has a de Sitter minimum at $ \vp=\vt=0$ where $V_{\rm min} =\Lambda$. Supersymmetry is broken at this minimum with $D_S W=M$,   the mass of gravitino is $m^2_{3/2}= {M^2\over 3} (1-{\Lambda\over M^2})$. The inflationary de Sitter valleys have a nice feature known for models with Minkowski minimum with unbroken SUSY, studied in \cite{Carrasco:2015rva}. These  valleys provide nice initial conditions for the inflation to start in these models.}
\label{Planck}
\end{center}
\vspace{-0.5cm}
\end{figure}
Here we will present an example where in disk variables the superpotential is relatively simple whereas the potential is not simple but satisfactory for our purpose. We take the inflaton shift-symmetric \K\, potential and the superpotential of the form
\be
K= -{3\over 2}   \alpha \log \left[{(1- Z\bar Z)^2\over (1-Z^2) (1-\overline Z^2)}  \right] +S\bar S\, , \quad S^2(x, \theta)=0\, ,  \quad W= \Big (S+ {1-Z^2\over b}\Big ) ( {\sqrt 3\alpha } \, m^2 \, Z^2 + M)\,   \ .
\label{KdiskExample}\ee

The same model in Killing variables $\Phi$,  where $Z= \tanh {\Phi\over \sqrt{6 \alpha}}$,  is
\be
 K= -3\alpha \log \Big [\cosh {\Phi-\bar \Phi \over \sqrt{6\alpha}} \Big] + S \bar{S} , \quad   W= \Big( {1\over b}{ \cosh ^{-2} \Big ({\Phi\over \sqrt {6\alpha}}\Big ) } + S\Big )  \Big ( \sqrt { 3 \alpha} \, m^2 \tanh^2 \Big ({\Phi\over \sqrt {6\alpha}}\Big ) +M \Big) .
 \label{newRexample} \ee
The potential at $S=0$ and $\vt=0$ has the form $
V_{\rm total}=  2 |g'(\vp)|^2 - 3  |g(\vp)|^2 +  |f(\vp)|^2
$, where in our case
\be
g(\vp)={1\over b}{ \cosh ^{-2} \Big ({\Phi\over \sqrt {6\alpha}}\Big ) }  \Big ( \sqrt { 3 \alpha} \, m^2 \tanh^2 \Big ({\Phi\over \sqrt {6\alpha}}\Big ) +M \Big)  , \quad  f (\vp) =    \sqrt { 3 \alpha} \, m^2 \tanh^2 \Big ({\Phi\over \sqrt {6\alpha}}\Big ) +M \ .
\ee
We have checked that the mass of the field $\vt$ is positive everywhere for all $\alpha >0.02$ and that during inflation the ratio ${m_{\vt}^2\over H^2}=6$. This can be also seen from the Fig. \ref{Planck}, where we plotted our potential. The inflationary de Sitter valleys are of the same width everywhere for larger and larger values of $\vp$. 

\begin{figure}[htb]
%\vspace*{-2mm}%\hspace{-3mm}
\begin{center}
\includegraphics[width=8cm]{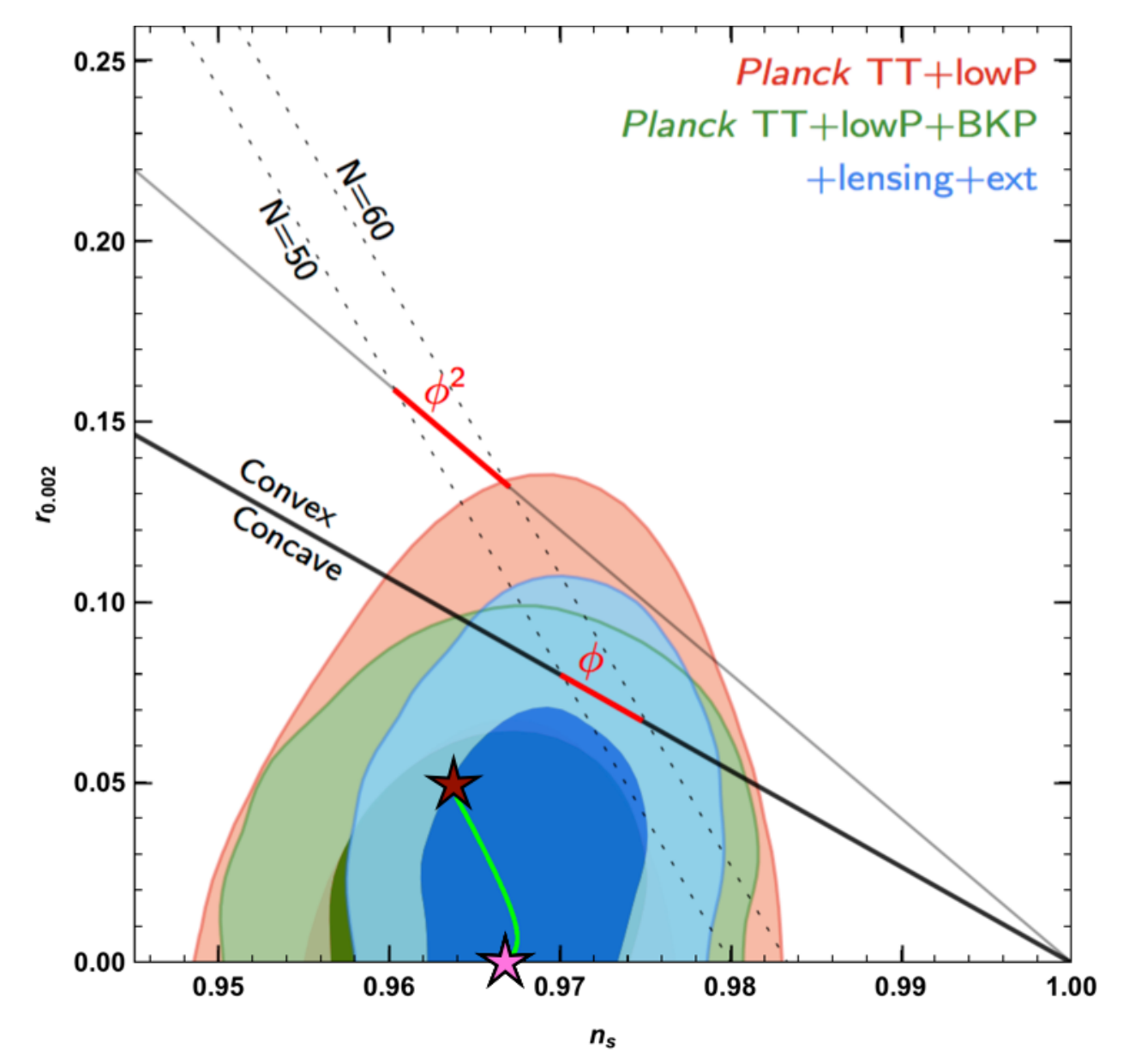}
\vspace*{-0.2cm}
\caption{\footnotesize  Predictions of the model \ref{KdiskExample}  for $20> \alpha > 1/3$ are shown by the thin green line. The top of the line indicated by the dark red star corresponds to $\alpha = 20$. The line ends at the pink star corresponding to $\alpha = 1/3$.
}
\label{flast}
\end{center}
%\vspace{-0.6cm}
\end{figure}

The predictions of this class of models for $n_{s}$ and $r$   practically coincide with the predictions of the models discussed in Sections 4.1 and 4.2 for $\alpha = O(1)$. However, at $\alpha \gg 1$ the predictions are somewhat different. We show these predictions  in Fig. \ref{flast} by a thin green line for $20> \alpha > 1/3$  and for the number of e-foldings $N = 60$. The top of the line indicated by the dark red star corresponds to $\alpha = 20$. The line ends at the pink star corresponding to $\alpha = 1/3$. We see that the predictions of this model  in the large interval  $20> \alpha > 1/3$ belong to the dark blue region favored by the Planck data.

Thus in the last two sections we have presented  several supergravity models where  ${\delta\rho\over \rho}$, $n_s$  and $\Lambda $ take their known observable values, whereas the gravitino mass $m_{3/2}$ and the tensor-to-scalar ratio $r$ are free parameters which can take a broad range of values.

\section{Discussion}
\label{sect:Discussion}

In this paper we have pursued a program of describing the main features of the universe evolution, early universe inflation and current acceleration compatible with the data,  as well as providing an explanation of the possible origin of the supersymmetry breaking and the mass of gravitino, compatible with the future data from particle physics. Certain features of our  four-parameter `primordial' supergravity models are motivated by the non-perturbative string theory. The origin of the nilpotent superfield $S^2(x, \theta)=0$ in these constructions is related to the D-brane physics, where one finds the fermionic Volkov-Akulov  goldstino multiplet \cite{Volkov:1973ix,rocek,Komargodski:2009rz} on the world-volume of the D-branes \cite{Kallosh:2014wsa}. 

Our new cosmological models in string theory inspired supergravity suggest a possible bottom-up ${\cal N}=1$ supergravity models of inflation which might lead to a successful phenomenology of the early universe and the one which is  accelerating now. These models also address the supersymmetry breaking issues. They  differ from more traditional string cosmology models which were developed during the last decade, the latest models being discussed in 
\cite{Flauger:2014ana,Buchmuller:2015oma,Dudas:2015lga} and other papers.
Our models have fundamental  connections to string theory via the nilpotent superfield associated with the fermions on the D-branes. Another connection is via  logarithmic \K\, potentials which are required for ${\cal N}\geq 2$ supergravity and are present in string theory motivated supergravity. And finally, the value of the positive cosmological constant in our models can be only explained with the reference to the string landscape.

The mass of gravitino,   $m_{3/2}$,  and the level of gravity waves, $r$,  are free parameters  in our new cosmological models, to be determined by the future experiments.  The progress in this direction was based on a better understanding of moduli stabilization and on the use of supergravity models with the universal spontaneous supersymmetry breaking via a fermionic goldstino multiplet. The reason for such universality is the following:  the nilpotency condition $S^2(x, \theta)=0$  for $S=s+ \sqrt{2}\, \theta\, \psi_s + \, \theta^2 F_s$ can be satisfied only if $F_s \neq 0$. In such case the sgoldstino is not a fundamental scalar anymore but is given by a bilinear combination of fermionic goldstino's divided by the value of the auxiliary field $F_s$
\be
 s= {\psi_s \psi_s\over 2 F_s} \ .
\ee 
There is no non-trivial solution if SUSY is unbroken and $F_s=0$, i.e. only $s=\psi_S=0$ solve  the equation $S^2(x, \theta)=0$. Thus by requiring to have  a fermion Volkov-Akulov goldstino nilpotent multiplet in supergravity theory we end up with the universal value of the supergravity potential at its minimum, with $e^K  |F_S |^2=M^2$
\be
V=e^K( |F_S |^2 - 3 |W|^2)= M^2- 3 m_{3/2}^2= \Lambda >0 \ .
\ee
{\it The new always positive  goldstino contribution originates in an updated version of the KKLT uplifting via the the $\overline D$3 brane, with manifest spontaneously broken supersymmetry} \cite{Kallosh:2014wsa}.  

 Our minimal supergravity models depend on two superfields,  one of them typically represented using either a Poincare-disk variable $Z$, or a half-plane variable $T$. A new variable $\Phi$ which we used extensively in this paper  describes the same geometry but in a Killing adapted frame where the metric does not depend on the inflaton direction. We call $\Phi$ a Killing variable.  We explained the relation between these three holomorphic variables in Section~\ref{sect:Killing}. The canonically normalized inflaton in our models is $\vp = {\rm Re} \, \Phi$. The inflaton partner scalar $\vt = {\rm Im} \, \Phi$  is supposed to vanish, which happens automatically during inflation in the models considered in this paper.  In all our models in $\Phi$-variables the inflaton shift-symmetric \K\, potential is $ K= -3\alpha\, \ln \Big [\cosh {\Phi-\bar \Phi \over \sqrt{6\alpha}} \Big] + S \bar{S}$
and the superpotential is $W= g(\Phi) + S f(\Phi)$. The nilpotent  multiplet $S$ does not have fundamental scalars, it has only a fermionic goldstino. 

In models with a canonical \K\, potential for the nilpotent multiplet $ K=  S \bar{S}$ stabilization of $\vt$ in all models presented in this paper does not require any additional stabilization terms,  as long as $\alpha  > 0.2$. For smaller $\alpha$ one can stabilize $\vt$ by adding a bisectional curvature term to the \K\, potential of the form (in disk variables)  $A(Z, \bar Z) S\bar S (Z-\bar Z)^2$. Thus, in presence of the nilpotent superfield $S$ the problem of stabilization of the direction orthogonal to the inflaton is solved during inflation as well as at the minimum of the potential.

An unexpected benefit from the new tools for moduli stabilization during inflation was realized very recently. Many examples of previously known supergravity models, compatible with current and future cosmological observations, can now easily describe dark energy via tiny de Sitter vacua,  and spontaneous breaking of supersymmetry. In this paper we provide examples of such generalizations of $\alpha$-attractor models \cite{Kallosh:2013hoa,Ferrara:2013rsa,Kallosh:2013yoa,Cecotti:2014ipa,Kallosh:2013tua,Galante:2014ifa,Kallosh:2015lwa,Kallosh:2015zsa,Carrasco:2015uma}. 
These models  interpolate between various polynomial models $\vp^{2m}$  at very large $\alpha$ and attractor line for $\alpha\leq 1$, see Figs. 1, 2. Therefore they are flexible with regard to data on B-modes, $r$. They provide a seamless natural fit to Planck data.
 For  these kinds of cosmological models we have shown that it is possible to break supersymmetry without an additional hidden sector, with a controllable  parameter of  supersymmetry breaking. With inflationary scale $\sim 10^{-5} M_{p}$ the scale of supersymmetry breaking can be $M\sim (10 ^{-13}- 10^{-14}) M_{p} $, compatible with the discovery of supersymmetry at LHC.  With $ M \gg 100- 1000$  TeV we will have equally good inflationary models, compatible with an absence of observed supersymmetry at LHC. In fact, such inflationary models are even easier to construct.
 
In this paper we developed two methods of constructing inflationary models with supersymmetry breaking and de Sitter minimum. One is the reconstruction method in 
Section~\ref{sect:Reconstruction}, which allows to take any desirable inflationary models, in particular our $\alpha$-attractor models, and enhance them by SUSY breaking and a small cosmological constant. 
An advantage of this method, following \cite{Dall'Agata:2014oka,Kallosh:2014hxa,Lahanas:2015jwa,Kallosh:2015lwa}, is that it is powerful and easy. It requires only a simple integration of a given function. Thus one can obtain nearly arbitrary inflationary potentials, just as it was done in \cite{Kallosh:2010ug,Kallosh:2010xz}, so one can fit any set of observational data in the context of supergravity-based models of inflation. Moreover, in all of these models one can introduce SUSY breaking of any magnitude without introducing extra scalars such as Polonyi field. It can be done while preserving all desirable inflationary predictions. Thus from the purely phenomenological point of view, the reconstruction method is a great tool offering us enormous flexibility.

On the other hand, this method does not use specific advantages of the cosmological attractors, including their geometric origin and stability of their predictions with respect to the change of the inflationary potential. In this sense, the method used for deriving the model described in  Section~\ref{sect:genModels}, as well as of some other similar models found earlier in \cite{Kallosh:2015lwa}, preserves the attractor features of the theory by construction, for all values of the SUSY breaking parameters and arbitrary cosmological constant.
Some of the features of these models (the existence of a dS valley of a constant width and depth shown in Fig. 3) play an important role in solving the initial conditions problem for inflation in these models. The details of this analysis can be found in \cite{Carrasco:2015rva} for $\alpha$-attractor models with a Minkowski minimum and unbroken SUSY.  Here we see that in generic models with de Sitter exit and controllable SUSY breaking, initial condition problem for inflation is solved just as in the simpler case studied in  \cite{Carrasco:2015rva}.

\section*{Acknowledgments}

We are grateful  to   S. Dimopoulos, M. Dine, S. Kachru, J. March-Russell, D. Roest, M. Scalisi, E. Silverstein, F. Quevedo and F. Zwirner  for a discussion of cosmology and particle physics related issues.  This work
was supported by the SITP and by the NSF Grant PHY-1316699.  RK is also supported by the Templeton foundation grant `Quantum Gravity Frontiers,' and AL is also supported by the Templeton foundation grant `Inflation, the Multiverse, and Holography.'   JJMC received support from the Templeton foundation grant `Quantum Gravity Frontiers,' and is supported by the European Research Council under ERC-STG-639729, `Predictive Quantum Field Theory'.

\end{document}